\documentclass[12pt]{article}
\newcommand{\sss}{\setcounter{equation}{0}}

\newtheorem{theorem}{THEOREM}[section]

\newtheorem{lemma}[theorem]{LEMMA}

\newtheorem{corollary}[theorem]{COROLLARY}

\newtheorem{definition}[theorem]{DEFINITION}

\def\ER{{\bf R}}
\def\beq{\begin{equation}}
\def\ene{\end{equation}}
\def\p{{\bf p}}
\def\v{{\bf v}}
\def\uv{\hat{\bf v}}
\def\bull{\begin{flushright} \vrule height 6pt width 6pt depth -.pt
\end{flushright}}
\def\l{L^2\left( \ER^n\right)}
\def \a{W_{\pm}^{\ast}}
\def\w{W_{\pm}}
\def \a{W_{\pm}^{\ast}}
\def\0{\{0\}}

\def\h{W_{1,2}}
\def\X{W_{1,p+1}}

\begin{document}

\baselineskip=23.6pt
\title{The Time-Dependent Approach to Inverse Scattering\thanks{{\sc ams}  
classification 35P25, 35Q40, 35R30 and 81U40. Research partially supported by 
Proyecto PAPIIT IN 105799. DGAPA-UNAM}}
\author{Ricardo Weder\thanks{Fellow Sistema Nacional de Investigadores}\\
Instituto de Investigaciones en Matem\'aticas Aplicadas y en Sistemas,\\
Universidad Nacional Aut\'onoma de M\'exico, \\Apartado Postal 20-726, M\'exico D.F. 
01000\\E-Mail: weder@servidor.unam.mx\\}
\date{}
\maketitle
\begin{center}

\vspace{2cm}
Notes of the lectures given at the Pan-American Advanced Studies Institute (PASI)
at the Mathematical Sciences Research Institute, Berkeley. October 29-November 2,
2001

\vspace{2cm}

\begin{minipage}{5.75in}
\centerline{{\bf Abstract}}\bigskip
In these lectures I  give an introduction to the time-dependent approach to inverse scattering, that has been developed recently.
The aim of this approach is to solve various inverse scattering problems with time-dependent methods that closely follow the physical (and geometrical) 
intuition of the scattering phenomena. This method has been applied to many linear 
and nonlinear scattering problems.
We first discuss the case of quantum mechanical potential scattering.
We  give explicit limits  for the high-energy behaviour of the scattering
operator that offer us formulae for the unique reconstruction of the potential.

Then, we  consider the case of the Aharonov-Bohm effect ( Schr\"{o}dinger operators with
singular  magnetic potentials and exterior domains). This is a particularly interesting inverse 
scattering problem that shows that in quantum mechanics  a magnetic field acts on a 
charged  particle -by means of the magnetic potential- even in regions where it is 
identically zero.

The key issue for these two problems is that at high-energies translation of 
the wave packet dominates over spreading during the interaction time.
\end{minipage}
\newpage
\begin{minipage}{5.75in}

In fact, in this limit it is sufficient for the calculation of the scattering 
operator to consider translation of wave packets rather than
their correct free evolution.

Finally, we study the nonlinear Schr\"{o}dinger equation with a potential. In 
this case, from the scattering operator we uniquely reconstruct the potential and 
the nonlinearity. For this purpose, we observe that in the small amplitude limit
the nonlinear effects become negligible and scattering is dominated by the linear 
term. Using this idea we prove that the derivative at zero of the nonlinear 
scattering operator is the linear one. With the aid of  this fact we first uniquely 
reconstruct the potential from the associated linear inverse scattering problem and 
in a second step we uniquely reconstruct the nonlinearity.
\end{minipage}
\end{center}

\section{Potential Scattering}\sss
First we briefly discuss time-dependent direct scattering theory in the 
particular case of potential scattering for the Schr\"{o}dinger equation.
It is important to keep in mind that physical scattering is a time-dependent 
phenomenon 
that studies the interaction of 
a finite-energy wave packet with a target. Initially, for large negative times,
the wave packet is far from the target and since the interaction is very small
its evolution is well approximated by an {\it incoming asymptotic state},
 $\phi_-$, that propagates according to the free dynamics, with the interaction  
set to zero. During the 
interaction time the wave packet is close to the target and, as the interaction is
strong,  the evolution of the wave packet is given by an 
{\it interacting state} that evolves according to the interacting dynamics. 
 But eventually, the wave packet flies away from the target and its evolution 
is again well approximated by an {\it outgoing asymptotic state}, $\phi_+$, that 
evolves according to the free dynamics. The scattering operator,\, $S$,\, is the 
operator that 
sends $\phi_-$ to $\phi_+$. The aim of scattering experiments is  to measure 
the transition probabilities, $(S \phi, \psi)$. 

One objective of the time-dependent 
approach to inverse scattering theory is to use in an essential way the physical 
propagation aspects to solve the inverse scattering problem and to obtain 
mathematical proofs that closely follow physical intuition. It is hoped that a good
physical understanding of the inversion mechanisms will be reflected in more 
transparent mathematical methods. In the stationary (frequency domain) method the 
physical solution is 
idealized as a time-periodic solution with infinite energy. By doing so, the 
propagation aspects of physical scattering are lost. This loss of physics 
 is then reflected in mathematical methods that do not give much 
information about the physics of the inversion. Moreover, in the stationary method
the wave packets (finite-energy solutions) are obtained using a generalized 
Fourier  transfom that integrates over the infinite-energy, time-periodic, solutions. 
Here the linearity of the direct scattering problem plays an essential role, 
because it is only 
in this case that linear combinations of solutions are solutions. On the contrary, 
the time-dependent approach does not use the linearity of the 
direct scattering problem in an essential way  and, as we will see below,
it has a natural extension to the case where the direct scattering problem is 
nonlinear.

But, let us be more specific.
We consider a quantum-mechanical particle in $\ER^n$ whose dynamics is 
described by the Schr\"{o}dinger equation,

\beq
i \frac{\partial}{\partial t} \Phi (t,x) =\frac{\p^2}{2m} \Phi (t,x)+ V(x) 
\Phi (t,x), \,\,\Phi (0,x)= \Phi_0(x)\in \l,
\label{1.1}
\ene
where $ t \in \ER, \,\,x \in \ER^n, n=2,\cdots$ , and $\Phi$ is complex valued,
$\p := -i \nabla$ is the momentum operator, and $ m >0$ is the mass of the particle. 
We take 
Planck's constant equal to one. The target is given by the potential, $V$, that is a 
real-valued function. For simplicity 
of the presentation, we consider the following class of bounded continuous 
short-range 
potentials. For the general case including singular and long-range potentials 
see \cite{ew2} and \cite{ew3}.

\beq
{\cal V}_{SR}:= \left\{ V \in C\left( \ER^n\right): 
\hbox{sup}_{|x| \geq R}\, \left| V(x)\right| \in L^1\left( [0, \infty )\right) 
\right\}.
\label{1.2}
\ene
 Both the free Hamiltonian, $H_0:= \frac{\p^2}{2m}$ and the interacting Hamiltonian, 
$H:= H_0 +V$ are self-adjoint in $\l$ with domain the Sobolev space,
$W_{2,2}$ \cite{8}. In what follows we denote by $\left\| \cdot \right\|$ the norm in $\l$. 
As is well known (see \cite{rs} for a general reference in scattering theory) if 
$V \in {\cal V}_{SR}$ -and also under much more general conditions- the wave operators,
\beq
\w := \hbox{s -} \lim_{t \rightarrow \pm \infty} e^{i t H}\, e^{-it H_0},
\label{1.3}
\ene 
exist and are complete. That is to say, the strong limits in (\ref{1.3}) exist and 
their range 
is equal to the subspace of absolute continuity of
$H$, ${\cal H}_{ac}$. Moreover, $\w$ are unitary from $L^2\left(\ER^n \right)$ onto 
$ {\cal H}_{ac}$. These facts give a mathematical basis to the  physical
description of scattering given above, as we explain now. 
The solutions to the free Schr\"{o}dinger equation 
(\ref{1.1}) with $V=0$ are given by $ e^{-itH_0}\Phi_-, \,\,\Phi_- \in \l$. 
They correspond 
to free  particles that travel in space without being perturbed by a potential. 
Since the potential is localized near zero an incoming
 particle that is localized at spatial infinity for very large negative times 
will 
be -to a good approximation- a  solution to the free Schr\"{o}dinger 
equation, $e^{-itH_0}\Phi_- $, with   
{ \it incoming asymptotic configuration }  $\Phi_-$. 
 As time goes by, and the incoming solution is near the scattering center (zero) 
it will 
feel the 
influence of the potential and we expect that it will be close to a solution
 of the interacting Schr\"{o}dinger equation (\ref{1.1}) with the potential,
$e^{-itH} \Psi$, with {\it interacting state}, $\Psi$ . But (\ref{1.3}) 
with $t \rightarrow -\infty$ tells us 
that there is a unique 
$\Psi:= W_- \Phi_- \in {\cal H}_{ac}$ such that this is true,

\beq
\lim_{t \rightarrow - \infty} \left\| e^{-itH} \Psi -e^{-itH_0} \Phi_- 
\right\|=0.
\label{1.4}
\ene
 
Furthermore, we expect that at later times the particle will escape the influence 
of the potential and that as $ t \rightarrow \infty$ it will travel to 
spatial infinity  where again it will be close to an outgoing solution of the 
free Schr\"{o}dinger  equation. In fact 
(\ref{1.3}) with $t \rightarrow \infty$ tells us that this is true
and that there is a unique {\it outgoing asymptotic configuration} 
$\Phi_+:= W_+^{\ast}\Psi$ 
such that,
\beq
\lim_{t \rightarrow  \infty}\left\| e^{-itH} \Psi - e^{-itH_0 }\Phi_+ 
\right\|=0.
\label{1.5}
\ene
In a scattering experiment, given  the {\it incoming free solution} one 
seeks to obtain information about  the {\it outgoing free solution}. This is actually parametrized by the 
corresponding Cauchy data at $t=0$, $\Phi_{\mp}$. The scattering operator is 
then, the operator that  assigns $\Phi_+$ to $\Phi_-$:

\beq
S:= W_+^{\ast}\, W_-.
\label{1.6}
\ene  
 We prove below that the 
high-energy limit of the scattering operator gives the Radon (or X-ray) transform 
of the potential. Inverting this transform we uniquely reconstruct the potential.  
 The mathematical proof closely follows physical intuition. The key issue is that at 
high energies {\it translation of the wave packets}  dominates over {\it spreading}
during the interaction time. In fact, in the high-energy limit  it is sufficient
for the calculation of the scattering operator to consider {\it translation of wave 
packets} rather than their correct free evolution. Since on this limit
 {\it spreading} occurs only when and  where the interaction is negligible, i.e. when the 
free and the interacting time evolutions are almost the same, the effect of 
{\it spreading} does not appear on the scattering operator. For this reason 
scattering
simplifies on the high-energy limit and we can uniquely reconstruct the potential.
We also obtain error bounds.
Let us consider  states, $\Phi_0$, with compact 
momentum support on the open ball, $B_{m \eta}$ of radius $ m \eta$ and center zero, 
\beq
\hat{\Phi_0} \in C^{\infty}_0 \left( B_{m \eta}(0)\right),
\label{1.7}
\ene
where $\hat{\Phi_0}$ denotes the Fourier transform of $\Phi_0$. The boosted  state,
\beq
\Phi_{\bf v}:= e^{im {\bf v}\cdot x} \Phi_0 \,\leftrightarrow \, 
\hat{\Phi}_{\bf v}= \hat{\Phi_0}(p-m -\v)\in C^{\infty}_0\left( B_{m \eta}(m {\bf v})\right),
\label{1.8}
\ene
has velocity support of radius $\eta$ around ${\bf v}$. Above we denote by
$B_{m \eta}(m \v)$ the open ball of center $m \v$ and radius $m \eta$. In the theorem below we 
use the 
high-velocity limit in an arbitrary fixed direction  $\hat{\bf v}:=
 \frac{\bf v}{|{\bf v}|},\,\,\, v:= |{\bf v}| \rightarrow \infty$. 
   
\begin{theorem}
Suppose that $V \in {\cal V}_{S R}$ and that for some $ 0 \leq \rho \leq 1$,

\beq
 (1+R)^{\rho} \hbox{\rm sup}_{|x| \geq R}\, \, \left| V(x)\right| \in 
L^1\left( [0, \infty )\right).
\label{1.9}
\ene 
Then, for  all $\Phi_{\v}, \Psi_{\v}$ as defined in (\ref{1.8})

$$
 iv \left( \left(S-I\right)\Phi_{\v},  \Psi_{\v} \right)
\equiv iv \left( e^{-im\v \cdot x}\,\left(S-I\right)\, e^{im\v \cdot x}\Phi_{0},  
\Psi_{0} \right)
$$
\beq
= \left(  \int_{-\infty}^{\infty} \, d \tau  V(x+ \hat{\v} \tau )      \Phi_0, 
\Psi_0 \right)+\cases{ o\left( v^{-\rho} \right),& $ 0 \leq  \rho < 1$,\cr\cr
 O\left( v^{-1 } \right),&  $\rho =1$.}
\label{1.10}
\ene
Moreover, the scattering operator, $S$, determines uniquely  the potential
$V \in {\cal V}_{S R}$.
\end{theorem}
\bull

Theorem 1.1 is proven in \cite{ew1} and \cite{ew3}. We give an idea of the proof.
By Duhamel's formula,

\beq
\w = I + i\int_0^{\pm \infty} dt \,e^{itH} V e^{-itH_0},
\label{1.11}
\ene

Then,
\beq
W_+ - W_- =i \int_{-\infty}^{\infty} dt\, e^{iHt}\, V\, e^{-itH_0}.
\label{1.12}
\ene
Since the  wave operators $\w$ are unitary,  $\a \, \w =I$, and moreover, as they satisfy 
the intertwinig relations,  
$ \w H_0 = H \w$, it follows from (\ref{1.6}) and (\ref{1.12}) that
$$
i (S-I)= i \left( W_+^{\ast} W_- - W_-^{\ast} W_-\right) = i 
\left( W_+ -W_- \right)^{\ast}
W_-
=  \int dt\, e^{iH_0t}\, V\, e^{-iHt}\, W_-
$$
\beq
=  \int dt\, e^{iH_0t}\, V\, W_-\, 
e^{-iH_0t}.
\label{1.13}
\ene
Taking a boost with velocity $\v$ and making the substitution $\tau =v t$, we 
obtain that,

\beq
e^{im \v \cdot x}iv (S-I) e^{-im\v\cdot x}= L_{\v}+R_{\v},
\label{1.14}
\ene
where,
\beq
L_{\v}:=\int \left[e^{-im\v\cdot x}\, e^{iH_0 \tau / v}  e^{im\v\cdot x}\right]\,V(x)\,  
\left[ e^{-im\v\cdot x}\, 
 e^{-i H_0 \tau /v}  e^{im\v\cdot x}\right]  \, d \tau,
\label{1.15}
\ene
and
\beq
 R_{\v}:= \int d \tau\, e^{-im\v\cdot x}\, e^{iH_0 \tau / v}  
\, V\, (W_- -I)
\, e^{-iH_0 \tau / v}  
e^{im\v\cdot x}.
\label{1.16}
\ene
$L_{\v}$ - that is the first Born approximation-
is the leading term which tends to a finite limit  if the velocity $v$ goes to 
infinity. We will use it to reconstruct the potential. The remainder, $R_{\v}$, 
represents multiple scattering. It is intuitively clear that each instance
of scattering by a short-range potential yields a factor $v^{-1}$: the strength
of the interaction as measured by $(S-I)$ or $(\w -I)$ is proportional to the time
which the particle spends in the interacting region  where the potential is strong.
This time is inversely proportional to the speed $v$ and since we have rescaled
multiplying  by $v$, we expect $R_{\v}$ to decay as $1/v$,  and that we can  
neglect it  as $v \rightarrow \infty$.

For any Borel function $f$
let us define the operator $f(\p)$ by functional calculus, or equivalently,  as 
$f(\p):= {\cal F}^{-1} f(\cdot ) {\cal F}$, where  ${\cal F}$ 
denotes the Fourier transform.

Under translation in momentum or configuration space, generated by $x$ or
$\p:= -i \nabla $, respectively, we obtain
\beq
e^{-im \v \cdot x} f(\p) e^{im\v \cdot x} = f(\p +m \v),
\label{1.17}
\ene
in particular,

\beq
e^{-im \v \cdot x} e^{-iH_0 \tau /v} e^{im\v\cdot x}= e^{-i \p \cdot \uv \tau}\,  
 e^{-i H_0 \tau /v}   e^{-im v  \tau/2}.
\label{1.18}
\ene
Moreover,
\beq
e^{i\p \cdot \uv \tau   } f(x) e^{-i\p \cdot \uv \tau} = f(x + \uv \tau).
\label{1.19}
\ene 
By (\ref{1.18}) the boosted free evolution consists of the classical translation in 
configuration space, $e^{-i\p \cdot \uv \tau}$ that is independent of $v$,
the term, $ e^{-i H_0 \tau /v}$ -that is responsible for the spreading of the 
wave packet-  and an unimportant phase. In the limit when $v \rightarrow \infty$ with
$\tau$ fixed,  $ e^{-i H_0 \tau /v}$ dissapears and it follows that we can replace the boosted
free evolution by the classical translation $ e^{-i\p \cdot \uv \tau} \Phi=
\Phi (x- \uv \tau)$. The point is that the classical translation is independent of
$v$ and the term responsible for the spreading goes to zero like $1/v$ and then it 
is irrelevant in the high velocity limit.
Then, by (\ref{1.19}) the pointwise limit under the integral in (\ref{1.15})  gives us,

\beq
\lim_{v \rightarrow \infty} L_{\v}= \int e^{i\p \cdot \uv \tau} V(x) 
 e^{-i\p \cdot \uv \tau} \, d \tau = \int V(x + \uv \tau) \, d \tau.
\label{1.20}
\ene 
This is the the desired limit   in Theorem 1.1. To make this argument rigorous 
we have to prove that the integrals above exist when applied to states as in 
Theorem 1.1 and we need an integrable uniform  bound for all large enough $v$ that allows us to 
use  the dominated convergence theorem to exchange limit and integration. 
At the same time, we will also prove that the 
remainder, $R_{\v}$, goes to zero as $1/v$. 

We first prove a propagation estimate
that expresses in a convenient way the fact that the solutions to the free 
Schr\"{o}dinger equation have rapid decay away from the classically allowed region,
i.e. away from the region in space where a classical particle that travels in 
straight lines with constant velocity would be. In fact the result follows easily
from the standard stationary phase estimate (see the Corollary to Theorem XI.14 in
\cite{rs}). Let us   denote by $  F( x \in {\cal M})$\ the operator of 
multiplication by the characteristic function of ${\cal M}$.
We designate by $\hat{f}$ the Fourier transform of $f$,
$\hat{f}(p):= 1 /(2 \pi)^{n/2} \int e^{-ip\cdot x}\, f(x)\, dx$.

\begin{lemma}
For any $f \in C^{\infty}_0( B_{m \eta})$, for some $ \eta > 0$,  and any  $ l=1,2, \cdots $, there is a constant 
$C_l$ such that the following estimate holds:
\beq 
\left\| F( x \in \tilde{{\cal M}})\, e^{-it H_0}\, f\left( \frac{{\p} - m  {\v}}{v^{\rho}} \right) 
F( x \in {\cal M}) \right\| \leq C_l  (1+r v^{\rho}+ \eta \,v^{2 \rho}|t|)^{-l},
\label{1.21}
\ene
for every $ {\bf v} \in \ER^n, \, t \in \ER , \,\,v > 0, \,\,\rho \in \ER$, and any measurable sets  
${\cal M}$ and ${\tilde{\cal  M}}$  in $\ER^n$
such  that
 $r:={\rm dist}\left( \tilde{{\cal M}}, {\cal M}+ {\bf v}t\right) -\eta v^{\rho } |t| \geq 0$.
\end{lemma}

\noindent {\it Proof:}\,\,by (\ref{1.17})-(\ref{1.19}) it is enough to prove the 
estimate,

\beq 
\left\| F( x \in \tilde{{\cal M}})\, e^{-it H_0}\, f\left( \frac{\p}{v^{\rho}} \right) 
F( x \in {\cal M}) \right\| \leq C_l \, (1+r v^{\rho}+ \eta \,v^{2 \rho}|t|)^{-l} ,
\label{1.22}
\ene
provided that, $ r:={\rm dist} \left( \tilde{{\cal M}}, {\cal M}\right)-\eta v^{\rho } |t| \geq 0$.
We have that,
\beq
\left\| F( x \in \tilde{{\cal M}})\, e^{-itH_0}\, f\left(\frac{\p}{v^{\rho}}\right) \, F(x\in {{\cal M}}) \phi \right\|^2
= \frac{1}{(2 \pi)^{n} } \int_{\tilde{{\cal M}}} dx \int_{{\cal M}} dy \int_{{\cal M}}
dz \,\tilde{f}_t (x-y)\,\phi (y)\, \overline{\tilde{f}_t (x-z) \,\phi(z)},
\label{1.23}
\ene 
where,

\beq
\tilde{f}_t (x):= \frac{1}{(2 \pi)^{n/2} } \int e^{ip\cdot x} e^{-i p^2 t/2m } \, f(p/ v^{\rho})
\, dp.
\label{1.24}
\ene

But, as $ 2 |\phi (y)|\,|\phi (z)|\leq |\phi (y)|^2 +|\phi (z)|^2 $,

\beq
\left\| F( x \in \tilde{{\cal M}})\, e^{-itH_0}\, f\left(\frac{\p}{v^{\rho}}\right) \, F(x\in {\cal M}) \phi \right\|^2
\leq  \frac{1}{(2 \pi)^{n}} \left[ \int_{|x|\geq r + \eta  v^{\rho} |t|} |\tilde{f}_t (x)|\, dx \right]^2
\, \|\phi \|^2.
\label{1.25}
\ene

Finally, by the Corollary to Theorem XI.14 in  \cite{rs}, for any  $N=1,2, \cdots ,$ there is a 
constant $C_N$ such that,

\beq
 |\tilde{f}_t (x)| \leq C_N \, v^{n \rho} \, (1+|x| v^{\rho}+  v^{2 \rho}|t|)^{-N},
\label{1.26}
\ene
for $|x| \geq \eta v^{\rho} |t|$. The lemma follows inserting (\ref{1.26}) in 
(\ref{1.25}).

\begin{corollary}
For any $ f \in C^{\infty}_0 \left( B_{m \eta}\right)$, any $v > 0$  and  for any  
$l=1,2, \cdots $, there is a 
constant $C_l$ such that,
\beq
 \left\| F(| x|\geq |\tau|/4 + \eta |\tau |/v )\, e^{-iH_0 \tau /v} \, 
f(p) 
F(| x |\leq |\tau |/ 8) \right\| \leq C_l (1+ |\tau |)^{-l}.
\label{1.27}
\ene
\end{corollary}

\noindent {\it Proof :}\,\, the Corollary follows from Lemma 1.2 with $ \rho =0$,
$\v =0$ and
$\tilde{{\cal M}}=\{ |x| > |\tau|/4 + \eta |\tau |/ v \}, \, 
{\cal M}= \{|x| \leq |\tau |/8 \}$. Observe that  
$r:={\rm dist}\left( \tilde{{\cal M}}, {\cal M} \right) - \eta | \tau |/ v \geq 
|\tau|/8  \geq 0$ .

\bull

By  (\ref{1.15})  and(\ref{1.18}) for any $\Phi_0$, as in (\ref{1.7}),

\beq
 L_{\v} \Phi_0= \int e^{iH_0 \tau / v} \,V(x+ \uv \tau )\,  
  e^{-i H_0 \tau /v} \, \Phi_0\, d \tau.
\label{1.28}
\ene

We  prove below the bound,

\beq
 \left\| e^{iH_0 \tau /v} V(x +\hat{\v} \tau ) e^{-iH_0 \tau /v}
\Phi_0  \right\| \leq h(|\tau |),
\label{1.29}
\ene
where $h$ is integrable. The left-hand side of (\ref{1.29}) side is dominated by
\beq
C \left\| V(x + \hat{\v} \tau ) F\left( |x| \leq \frac{|\tau |}{2}\right) 
\right\| + C \left\| F\left(|x| \geq \frac{|\tau|}{2}\right) e^{-iH_0
\tau /v} \Phi_0 \right\|.
\label{1.30}
\ene
The first summand is bounded by
$h_1(|\tau |):= \sup_{| y| \geq |\tau |/2} |V ( y)|$ which is independent
of $v$ and integrable by (\ref{1.2}). If $v \geq 4 \eta $ ( with $\eta $
the radius of the velocity support of $\Phi_0$), the second term describes
free propagation into the {\it classically forbidden region } and it is 
rapidly decaying. Let $g \in C^{\infty}_0 \left(B_{m \eta}\right)$ 
be such that
$g(\p )\Phi_0= \Phi_0$. Then, for $v \geq 4 \eta$,

$$
\left\| F\left(|x | \geq \frac{|\tau |}{2}\right) e^{-iH_0 \tau
/v} \Phi_0 \right\| \leq \left\| F\left( |x | \geq \frac{|\tau |}{8}\right) \Phi_0
\right\|
$$
\beq
 + C \left\| F\left(|x |\geq \frac{|\tau |}{4}+ \eta \frac{|\tau |}{v}
\right) e^{-iH_0 \tau /v} g(\p ) F\left( |x | 
\leq \frac{|\tau |}{8}\right) \right\|:= h_2 (\tau ) +h_3(\tau ).
\label{1.31}
\ene
The two terms on the right-hand side of (\ref{1.31}) have rapid decay as 
$|\tau| \rightarrow \infty$ uniformly on $v$, for $v \geq 4 \eta $ . In the case of the first
term this is obvious because $\Phi_0$ belongs to Schwartz space. For the 
second term it follows from Corollary 1.3. Defining, $h:= h_1+h_2+h_3$  we obtain 
the integrable bound (\ref{1.29}) for all $ v \geq 4 \eta$. By the dominated convergence 
theorem we can take pointwise limit under the integral in (\ref{1.28}) and we obtain,

\beq
\hbox{s -}\lim_{v \rightarrow \infty} L_{\v} \Phi_0 = 
 \int_{-\infty}^{\infty} \, d \tau  V(x+ \uv \tau ) \Phi_0.
\label{1.32}
\ene
We now  prove that the  remainder  $R_{\v}$ ({\it multiple scattering})
is one order  smaller, i.e., that it goes to zero as $1/v, \,\,v \rightarrow \infty$. 
By (\ref{1.11}), (\ref{1.16}) and (\ref{1.18}) have that,

$$
  \left|\left( R_{\v} \Phi_0, \Psi_0 \right)\right|
\leq \left|\int_{-\infty}^{\infty} d \tau \left(\int_{-\infty}^0 \, dt \,  e^{it H}\, V e^{-i(t+ \tau /v)H_0} e^{im \v \cdot x} 
\Phi_0, V e^{-iH_0 \tau /v} e^{im \v \cdot x} \Psi_0 \right) \right|  
$$
\beq \leq \frac{1}{v}
\left(\int_{-\infty}^{\infty} d\tau  \left\| V e^{-iH_0 \tau/v} e^{im \v \cdot x} \Phi_0 \right\|
 \right)\,
 \left( \int_{-\infty}^{\infty}d\tau \left\| V e^{-iH_0 \tau /v} 
e^{im \v \cdot x}\Psi_0\right\|\right).
\label{1.33}
\ene
 Using (\ref{1.17})-(\ref{1.19}) and (\ref{1.29}) we prove  that,
\beq
  \int_{-\infty}^{\infty}d\tau \left\| V(x )e^{-iH_0 \tau/v} e^{im \v \cdot x} 
\Psi_0\right\|=
  \int_{-\infty}^{\infty}d\tau  \left\| V(x +\uv \tau) e^{-iH_0 \tau /v} \Psi_{0}
\right\| \leq C
\label{1.34}
\ene
uniformly in $v$. By (\ref{1.33}) and (\ref{1.34})

\beq
\left|\left( R_{\v} \Phi_0, \Psi_0 \right)\right| \leq C / v, v \geq 4 \eta.
\label{1.35}
\ene

By (\ref{1.14}), (\ref{1.32}) and (\ref{1.35})
$$
\lim_{v \rightarrow \infty} iv \left( \left(S-I\right)\Phi_{\v},  \Psi_{\v} \right)
= \lim_{v \rightarrow \infty} iv \left( e^{-im\v \cdot x}\,\left(S-I\right)\, 
e^{im\v \cdot x}\Phi_{0},  
\Psi_{0} \right)
$$
\beq
= \left(  \int_{-\infty}^{\infty} \, d \tau  V(x+ \hat{\v} \tau )      \Phi_0, 
\Psi_0 \right).
\label{1.36}
\ene
This is the correct limit as required in Theorem 1.1. We estimate the error term in a
similar way. See \cite{ew3} for details.

Let us denote: $ x^{\perp}:= x -(x \cdot \hat{\v}) \hat{\v}\equiv
x -x^{\parallel} \hat{\v}$. Then, the integral

\beq
W(x^{\perp}; \hat{\v}):= \int d \tau V(x +\hat{\v} \tau),
\label{1.37}
\ene
exists and is continuous by (\ref{1.1}). The set of all $\Phi_0, 
\Psi_0$ is rich enough to determine for any $\hat{\v}$ the continuous function
$ W(\cdot , \hat{\v})$ from the r.h.s. of (\ref{1.10}). For $n=2$,  
$W(x^{\perp}; \hat{\v})$ is the Radon transform of the square integrable potential $V(x )$.
It is well known that the Radon transform determines $V$ uniquely (\cite{he}, p. 115). If $n \geq 3$
it is the $X$-ray transform. However, one can fix arbitrarily 
$(x_3,x_4, \cdots x_n )$ and apply the same to the resulting two-dimensional 
function. Then, varying $\hat{\v}$ in a plane is actually sufficient to reconstruct
$V(x)$ from  $W(x^{\perp}; \hat{\v})$. This completes the proof of Theorem 1.1.

The time-dependent approach is quite flexible. It has 
been applied to many  inverse scattering problems. In \cite{ew2}, \cite{ew3} 
to N-body systems with singular and long-range potentials, in \cite{we2} to 
the N-body Stark effect, and in \cite{ew4} to two-cluster scattering.
In \cite{we3} the case of N-body systems with time-dependent potentials was treated.
The case of regular magnetic fields on $\ER^n$ 
was considered in \cite{a1}, \cite{a2}, and \cite{a3}. The relativistic 
Schr\"{o}dinger operator, and the Dirac and Klein-Gordon equations were studied 
in \cite{j1} and \cite{j2}. In \cite{it} the Dirac equation with time-dependent electromagnetic
potentials was considered. For the case of the Aharonov-Bohm effect, see Section 2. 
For references  on the stationary theory see
\cite{we1} and \cite{ew3}. In all of these papers the direct scattering problem is 
linear. For the case when the direct scattering problem is nonlinear see Section 3.

\section{The Aharonov-Bohm Effect}\sss
We now discuss the Aharonov-Bohm effect \cite{ab}. Aharonov and Bohm 
considered the scattering of an electron off the magnetic
field of a tiny solenoid, idealized as having infinite length and zero radius 
(scattering off a thread of magnetic flux). We assume that the solenoid is located on the 
vertical axis of the coordinate system, and in consequence it is enough to consider scattering in the plane orthogonal to 
the solenoid, as the problem is invariant under translation along the vertical direction. 
The magnetic field of an unshielded solenoid at zero is given by, $ B_s \,\delta (x)
\linebreak $ with $ x=(x_1,x_2) \in  \ER^2$ (this 
actually corresponds to a magnetic field in  $\ER^3$ with components $(0,0, B_s \,\delta (x))$.
We also assume that there is a regular magnetic field $B_{R} \in C^1_0 (\ER^2)$ that is 
continuously differentiable and has compact support. The total magnetic
field is written as,
\beq
B:=  B_s \,\delta (x) + B_R,
\label{2.1}
\ene
 where, of course, $B_s$ is a real constant and $B_R$ is a real-valued function. In order to define the 
Schr\"{o}dinger operator for an electron in the presence of $B$ we have to introduce a magnetic
potential. The magnetic potential for  $B_s \,\delta (x)$ in the Coulomb 
gauge is given by,
\beq
A_0:= \frac{\alpha_{\0} }{ |x|^2} \left[\begin{array}{c} -x_2 \\ x_1 \end{array}
\right], 
\label{2.2}
\ene
where we denote $\alpha_{\0}:=\frac{B_s  }{2 \pi}$. Observe that  $\alpha_{\0}$ is the flux ( across zero) of the singular magnetic field normalized by $2\pi$. It is easily checked that 
$\nabla \times A_s:= \frac{\partial}{\partial x_1} A_{s,2}
-\frac{\partial}{\partial x_2} A_{s,1}= B_s \,\delta (x)$ and that $ \nabla \cdot A_s=0$,
with the derivatives taken in distribution sense in ${\cal D}'$.
The magnetic potential in the Coulomb gauge for $B_R$ is given by,
\beq
A_R= \frac{1}{2 \pi} \int B_R(x-y) \left[\begin{array}{c} -\hat{y}_2 \\ \hat{y}_1 
\end{array}\right]\, \frac{dy}{|y|},
\label{2.3}
\ene
where, $ \hat{y}:= \frac{y}{|y|}$. Clearly, $A_R\in C^1 \left(\ER^2, \ER^2\right)$.
The magnetic potential for $B$ in the Coulomb gauge is given by,
\beq
A_c:= A_s+ A_R.
\label{2.5}
\ene

As is well known, the magnetic potential is not uniquely defined by the magnetic field; there is
always the possibility of a gauge transformation. We introduce
below a general class of magnetic potentials that is convenient for our purposes.

\begin{definition}
We denote by ${\cal A}_{\0}(\alpha_{\0},B_R)$ the set of all real-valued 
$A \in C^1\left( \ER^2 \setminus 0, 
 \ER^2 \right) \cap L^1_{{\rm loc}} \left( \ER^2 , \ER^2 \right)$, with $ \nabla \times A =
 2 \pi \alpha_{\0} \,\delta (x) + B_R$, in  ${\cal D}'$. Moreover, we assume that
$A(x)=O\left(|x|^{-1}\right), |x| \rightarrow \infty$ and that,
\beq
a(r):= \sup_{|x| \geq r} |A(x)\cdot \hat{x}| \in L^1 ([0, \infty )).
\label{2.6}
\ene 
\end{definition}

Let us now study the case where a general singular magnetic field
is contained inside an infinite  cylinder, 
with axis along the vertical direction, and transversal section $K$, where $K$ is a compact    
subset of $ \ER^2 $.  The purpose of the cylinder is to shield the singular magnetic field from 
the incoming electrons. As we will see below, we cannot hope that the scattering operator
determines uniquely the magnetic field inside $K$. In fact, we can only determine the 
(normalized ) flux of the magnetic field across $K$ modulo 2. This suggests that instead of
specifying the magnetic field inside $K$ we only fix the magnetic flux across $K$, 
normalized by $2 \pi$, that by Stokes' theorem is given by the circulation of the magnetic potential, $A$, along $ \partial K$,

\beq  
 \alpha_K := \frac{1}{2 \pi} \int_{\partial K} A(x)\cdot dx,
\label{2.7}
\ene
where we integrate in counter-clockwise sense.
Of course, we also specify a regular magnetic field,
$B_R$,  outside of $K$.
The magnetic flux $\alpha_K$ could be produced, for example, by a finite number of
solenoids inside $K$, and also by a regular magnetic field contained inside
$K$, or by a combination of both. The considerations above suggest the definition of the following 
class of magnetic potentials. In what follows
we  denote,  $\Omega:= \ER^2 \setminus K$, where for the unshielded solenoid, $K=\{0\}$. 

\begin{definition}
Let $K$ be a compact set such that $0 \in K$ and that its boundary,
$\partial K$, is a simple, closed, $C^1$ curve.  Then, for any  $\alpha_K \in \ER$ and any real-valued 
$B_R \in C^1_0 \left( \overline{\Omega}\right)$  we denote by 
${\cal A}_K \left( \alpha_K, B_R\right)$  the set of all real-valued 
$A \in C^1\left( \overline{\Omega}, 
 \ER^2 \right)$ with $ \nabla \times A = B_R$ and such that,
\beq
\alpha_K =  \frac{1}{2 \pi}  \int_{\partial K} A(x)\cdot dx,
\label{2.8}
\ene
where we integrate in counter-clockwise sense.
 Moreover, we assume that
$A(x)=O\left(|x|^{-1}\right), |x| \rightarrow \infty$ and that,
\beq
a(r):= \sup_{x \in \Omega ,|x| \geq r} |A(x)\cdot \hat{x}| \in L^1 ([0, \infty )).
\label{2.9}
\ene 
\end{definition}
 
The formal Hamiltonian is the operator,
\beq
h_A:= \frac{\left( \p -A \right)^2}{2 m},
\label{2.10}
\ene
 with domain $C^2_0 \left( \Omega \right)$, $ \p := -i \nabla$ is the momentum 
operator, and  $ m> 0$ is the mass of the electron. We take Planck's constant  and the speed of light all equal to one and the  charge of the electron
equal to minus one.  
The quadratic form associated to $h_A$ is given by,
\beq
q_A( \phi ,\psi ):= \left( (\p -A) \phi , (\p -A) \psi \right),
\label{2.11}
\ene
with domain  $C^1_0 \left( \Omega \right)$. The form $q_A$ is non-negative and 
closable.
The Hamiltonian, $H_A$, is the self-adjoint operator in $L^2 ( \Omega)$ associated to the closure 
of $q_A$ (see \cite{ka}). $H_A$ is the extension of $h_A$ with Dirichlet boundary condition on 
$\partial \Omega$. 
Let $J$ be the identification operator from $L^2 \left( \ER^2 \right)$ onto   
$L^2 \left( \Omega \right)$ given by multiplication by the characteristic function of $\Omega$. In the case $K= \0$ we take $J=I$.
The unperturbed Hamiltonian is given by $H_0:= \frac{\p^2}{2m}$, with domain the Sobolev space
$W_{2,2}\left( \ER^2 \right)$. The wave operators are defined as,
\beq
\w (A):= \hbox{s -}\lim_{t \rightarrow \pm \infty} e^{it H_A}\, J\, e^{-it H_0}.
\label{2.12}
\ene
We prove in \cite{we11} that if $A \in {\cal A}_K (\alpha_K,B_R)$, 
the strong limits exist and are isometric.  The scattering operator is given by,

\beq
S(A):=  W_+^{\ast}\left( A \right)\,\,W_- \left( A \right).
\label{2.13}
\ene
Note that to define $H_A$ we only use  the values of $A$ in $\Omega$.
This means that as long as $A$ is fixed in $\Omega$, we can change the magnetic 
potential in the interior of $K$ without changing $H_A$. Note however that as 
$ A\in C^1 \left( \overline{\Omega}\right)$ the flux
$\alpha_K$ is uniquely defined by the values of $A$ in $\Omega$. 
This explains why we cannot hope to uniquely reconstruct the magnetic field 
inside $K$ from the scattering operator and makes it plausible that we can 
reconstruct $\alpha_K$. 

As we said above, the only purpose of the obstacle, $K$, is to shield the 
incoming electron from the magnetic
field, and in order to separate the scattering effect of the magnetic potential from that
of the obstacle, we consider asymptotic configurations that have negligible interaction
with $K$ for all times in the high-velocity limit. For this purpose, given $\hat{\v}\in S^1$,
let us denote,
\beq
 \Omega_{\uv}:= \left\{ x \in \Omega : x+ \hat{\v} \tau \in \Omega,\, \hbox{for all}\, \tau \in 
\ER \right\}.
\label{2.14}
\ene    
Given $\v \in \ER^2$ we take asymptotic configurations $\Phi \in C^{\infty}_0 ( \Omega_{\hat{\v}})$,
where $ \hat{\v}:= \v / v$, with $ v:= |\v|$. The  free evolution boosted by $\v$ is given by
 $ e^{-im \v \cdot x} \, e^{-itH_0} \, e^{im \v \cdot x}=  e^{-im v^2 t/2} 
e^{-i \p \cdot \v t} e^{-itH_0},$ and -to a good approximation- in the limit when $ v \rightarrow
\infty $ with $\hat{\v}$ fixed this can be replaced (modulo an 
unimportant phase) by the classical translation  $e^{-it\p \cdot \v}$. Then, in the 
high-velocity limit it is a good approximation to assume that the free evolution of our 
asymptotic configuration is given  by
$e^{-it\p \cdot \v} \Phi_0 = \Phi_0 (x- \v t) $, and as 
$\Phi_0 \in C^{\infty}_0 ( \Omega_{\hat{\v}})$, it has negligible interaction with $K$ 
for all times. In the following theorem we evaluate the high-velocity asymptotics of the 
scattering operator.

\begin{theorem}
Suppose that $A \in {\cal A}_K(\alpha_K, B_R)$ and that $\Phi_0, \Psi_0 \in C^{\infty}_0\left(\Omega_{\uv}
\right)$. Let 
$ \Phi_{\v}, \Psi_{\v}$ be the boosted asymptotic configurations,
\beq 
\Phi_{\v}:= e^{im \v \cdot x} \Phi_0, \,\, \Psi_{\v}:= e^{im \v \cdot x} \Psi_0.
\label{2.15}
\ene
Then, 

\beq
 \left( S(A) \Phi_{\v}, \Psi_{\v}\right)= \left( e^{i \int_{-\infty}^{\infty} \uv\cdot A(x+ 
\uv \tau )
\, d\tau} \Phi_0, \Psi_0 \right)+ O\left(\frac{1}{|v|}\right),\,\,\, |v| \rightarrow \infty.
\label{2.16}
\ene
\end{theorem}

We prove in \cite{we11} that (\ref{2.16}) determines uniquely the 
Radon transforms,
\beq
\int_{-\infty}^{\infty}B_R (x  + \uv \tau) \, d \tau, \,\,\, x \in \Omega_{\hat{v}}.
\label{2.17}
\ene
By the support theorem for the Radon transform  \cite{na},  $B_R$ is uniquely
determined in $\Omega$, provided that $K$ is convex. The fact   that 
the  magnetic flux, $\alpha_K$, is determined modulo 2 follows from an explicit 
calculation. This gives us 
the  following theorem.

\begin{theorem}
Suppose that $A^{(j)}\in {\cal A}_K\left(\alpha^{(j)}_K, B^{(j)}_R\right),
j=1,2$ and that $K$ is convex. Then, if $S\left(A^{(1)}\right)= S\left(A^{(2)}\right)$, we have that,
 $\alpha^{(1)}_K= \alpha^{(2)}_K$ modulo $2$ and that $B^{(1)}_R(x)=
B^{(2)}_R(x), x \in \Omega$. 
\end{theorem}
For a complete study of this problem including proofs and an analysis of gauge transformations see
\cite{we11}.
Observe that since (\ref{2.17}) is obtained from (\ref{2.16}) we actually only 
need to know the 
high-velocity limit of the scattering operator. Moreover, our proof gives a method 
for the 
reconstruction of $B_R$ in $\Omega$. Note that in spite of the fact that the 
scattering of the 
electron takes place outside of $K$, we determine the magnetic flux  in 
$K$- modulo 2- from the scattering 
operator. This is the Aharonov-Bohm effect \cite{ab} that shows that in quantum 
mechanics  the magnetic field acts on a charged particle -by means of the magnetic
 potential-
even in regions where it is identically zero.
Nicoleau has proven in \cite{ni} the following result using stationary methods. 
Suppose that   the magnetic field, $B$, is infinitely differentiable in $\ER^2$ 
and has 
compact support, and that   $K$ is compact, convex, $ 0 \in K$ and 
$\partial K$  is smooth. Then, if  $S\left(A^{(1)}\right)= S\left(A^{(2)}\right)$,
 with $A^{(j)},\,\, j=1,2$, the Coulomb gauge potentials,
then the two magnetic fluxes across $\ER^2$ are equal modulo $2$ and     
$B^{(1)}_R(x)=
B^{(2)}_R(x), x \in \Omega$.  The case of an unshielded solenoid with $K=\{0\}$ is 
not covered by the result of \cite{ni}. Note that this is actually the problem 
considered by  Aharonov and Bohm  \cite{ab}.
In the case where the interior of $K$ is non-empty, \cite{ni} considers the 
situation where the magnetic flux across $K$ is produced by a magnetic field in
$C^{\infty}_0 \left( \ER^2 \right)$. Our result is considerably more general
in the sense that we study the case where the magnetic flux across $K$ is produced
 by any magnetic field inside $K$. The only restriction is that the magnetic 
flux across $K$ has to be finite.
We could as well consider the case where there is also a scalar potential -as 
is done in \cite{ni}- however, we do not pursue that direction here.

\section{The Non-Linear Schr\"{o}dinger Equation} \sss

We discuss now the non-linear Schr\"{o}dinger equation with a potential on 
 the line. The goal is  to give a method to uniquely reconstruct the potential
and the nonlinearity. This problem was solved in  \cite{we4}, \cite{we5},
\cite{we6}, \cite{we7} and 
\cite{we8}. In these papers the multidimensional case was also considered.
For the  nonlinear Klein-Gordon equation see \cite{we12}, \cite{we9} and \cite{we10}.

Let us consider the following non-linear Schr\"{o}dinger equation with a potential
\beq 
i \frac{\partial }{\partial t}u(t,x)= -\frac{\partial^2}{\partial x^2} u(t,x)+V_0(x)u(t,x) +
F(x,u),\,\,u(0,x)= \phi (x),
\label{3.1}
\ene
where $t,x \in \ER$, the potential, $V_0$, is a real-valued function, and $F(x,u)$ is a 
complex-valued function.

Let $H_0$ denote the 
unique self-adjoint realization of $-\frac{d^2}{d x^2}$ with domain the Sobolev
space $W_{2,2}$ \cite{8}.
For any $\gamma \in \ER$,  $L^1_{\gamma}$ denotes the Banach space of all
complex-valued measurable functions, $\phi $, defined on $\ER$ and such that
\beq
\left\|\phi \right\|_{L^1_{\gamma}}:= \int |\phi (x)| \, (1+|x|)^{\gamma} \, dx
< \infty .
\label{3.4}
\ene
If $V_0 \in L^1_1$ the differential expression $\tau :=-\frac{d^2}{d x^2}+V_0(x)$ is 
essentially self-adjoint. 
We denote by $H$ the unique self-adjoint realization of $\tau$. The $H$ has a finite 
number of negative eigenvalues, it 
has no positive or zero eigenvalues, it has no singular-continuous spectrum and
 the absolutely-continuous spectrum is $[0, \infty )$ (for these results see \cite{wei}).
The wave operators are defined as in the multidimensional case,
\beq
W_{\pm}:= \hbox{s}-\lim_{t \rightarrow \pm \infty} e^{itH} e^{-itH_0}.
\label{3.4b}
\ene
The limits in (\ref{3.4b}) are taken in the strong topology in $L^2$.
As is  well known (see \cite{s}), the $W_{\pm}$ exist and are complete. 
The scattering operator for the linear Schr\"{o}dinger equation ((\ref{3.1}) with
$F=0$) is defined as follows:
\beq
S_L:=W_+^{\ast} \, W_-.
\label{3.7b}
\ene

The key issue for the scattering theory for  equation  (\ref{3.1}) is the
following time-dependent $L^p-L^{\acute{p}}$ estimate that we proved in  \cite{we5},
\begin{equation}
\left\|e^{-itH} P_c \right\|_{{\cal B}\left( L^p,L^{\acute{p}}\right)} \leq
\frac{C}{t^{(\frac{1}{p}-\frac{1}{2})}},\,  t > 0,
\label{p}
\end{equation}
for some constant $C,\,\, 1 \leq p  \leq 2$, and $\frac{1}{p}+\frac{1}{\acute{p}}=1$,
and where $P_c$ denotes the projector onto the space of continuity of $H$.
The $L^p-L^{\acute{p}}$ \,estimate \,(\ref{p}) expresses the dispersive nature of 
the solutions to the  linear 
Schr\"{o}dinger equation with initial data on the continuous subspace of
$H$. It gives a quantitative meaning to the 
{\it spreading of the wave packets}. In  typical applications the nonlinearity, F, is 
proportional to a high-enough power of $u$. This 
type of nonlinearity makes the solutions to (\ref{3.1}) even larger where they
are already large. On the other hand, the {\it spreading } of the associated 
linear equation prevents the solution from becoming
to large, provided that the initial data was small enough. It is the 
balance  from these two
phenomena that is at the heart of small-amplitude scattering theory. Eventually, the 
{\it spreading} prevents the solution from blowing up in a finite time, and for 
large times the evolution is dominated by the linear part in the sense that the 
solution is asymptotic to a solution of the  linearized equation. This is the 
physical content of Theorem 3.1 below. By the same argument, on the small amplitude 
limit
the nonlinear effects become negligible and scattering is dominated by the linear
term. This fact is expressed in a quantitave way by Theorem 3.2 that allows us to 
reconstruct the linear scattering operator from the derivative at zero of the 
nonlinear scattering operator. It is interesting to remark that the {\it spreading
of the wave packets} -that is irrelevant on the high-energy limit in the linear 
case-
is actually essential on the low-energy (small amplitude) limit 
in the non-linear case.
 
  Before we state our results we introduce some standard 
definitions and notations.
We say that $F(x,u)$ is a $C^k$ function of $u$ in the real sense if
for each fixed $x \in \ER$, 
$\Re F$ and $\Im F$ are $C^k$ functions of the real and
imaginary parts of $u$. We will   assume that $F$ is $C^2$ in the real sense and
that 
$\left(\frac{\partial}{\partial x} F\right)(x,u)$ is $C^1$ in the real sense. 
If $F=F_1+iF_2$ with $F_1, F_2$
real-valued, and $u=r+is,\,\, r,s \in \ER$ we denote,
\beq
F^{(2)}(x,u):=\sum_{j=1}^2\left[\, \left| \frac{\partial^2}{\partial r^2} F_j (x,u)
\right|+ \left| \frac{\partial^2}{\partial r  \partial s} F_j (x,u)\right|
+\left| \frac{\partial^2}{\partial s^2} F_j (x,u)\right|\,\right],
\label{3.2}
\ene
\beq
\left(\frac{\partial}{\partial x}F\right)^{(1)}(x,u):=\sum_{j=1}^2\left[\, \left| 
\frac{\partial}{\partial r} \left(\frac{\partial}{\partial x}F_j\right)(x,u)\right|
+\left| \frac{\partial}{\partial s} \left(\frac{\partial}{\partial x}F_j\right) 
(x,u)\right|\,\right].
\label{3.3}
\ene
For any pair $u,v$
of solutions to the stationary Schr\"{o}dinger equation:
\beq	
-\frac{d^2}{d x^2} u+V_0u = k^2 u,\, k \in {\bf C}^+,
\label{3.8}
\ene
let $[u,v]$ denote the Wronskian of $u,v$:
\beq
[u,v]:= \left(\frac{d}{d x}u\right) v- u \frac{d}{d x}v.
\label{3.9}
\ene
Let $f_1(x,k)\approx e^{ikx}, x \rightarrow \infty, f_2(x,k)\approx e^{-ikx}, x \rightarrow -
\infty$, be the Jost solutions to (\ref{3.8}) (see 
for example \cite{DT}). A potential $V_0$ is said
to be {\it generic} if $[f_1(x,0),f_2(x,0)]\neq 0$ and $V_0$ is said to be
{\it exceptional} if $[f_1(x,0),f_2(x,0)]=0$. If $V_0$ is {\it exceptional} there is 
a bounded solution to (\ref{3.8}) with $k^2=0$
(a half-bound state or a zero-energy resonance). The trivial 
potential, $V_0=0$, is {\it exceptional}. 
Let us denote,
$$
M:= \left\{ u \in C(\ER, W_{1,p+1}): \sup_{t \in \ER} (1+|t|)^d \|u\|_{W_{1,p+1}}
< \infty \right\}, 
$$
\beq
\hbox{with norm} \|u\|_M:= \sup_{t \in \ER}
(1+|t|)^d \|u\|_{W_{1,p+1}},
\label{3.10}
\ene
where $ p \geq 1$,\, and \, $d:= \frac{1}{2} \frac{p-1}{p+1}$. For functions 
$u(t,x)$ defined in $\ER^2$
we simply write $u(t)$, instead $u(t,\cdot)$. The $W_{k,p}$ are the Sobolev spaces 
\cite{8}. 
The small-amplitude scattering operator is given in 
the following theorem.
\begin{theorem}
Suppose that $V_0 \in L^1_{\gamma}$, where in the {\rm generic case} $ \gamma > 3/2$
and in the {\rm exceptional case} $ \gamma > 5/2$,  that $H$ has no negative 
eigenvalues, and that
\beq
N(V_0):= \sup_{x \in \ER} \int_x^{x+1} |V_0(y)|^2\, dy < \infty.
\label{3.10b}
\ene
 Furthermore, assume that $F$ is $C^2$ in the real sense, that 
$F(x,0)=0$, and that for  each fixed $x \in \ER$ all the first order derivatives,
 in the real sense, of $F$ vanish at zero. Moreover, suppose that 
$ \frac{\partial}{\partial x}F$ is $C^{1}$
in the real sense. We assume that the following estimates hold:
\beq
F^{(2)}(x,u)= O\left(|u|^{p-2}\right),\,\,\,
\left(\frac{\partial}{\partial x}F\right)^{(1)}(x,u)= O\left(|u|^{p-1}\right),\,
 \,u \rightarrow 0,
\label{3.11}
\ene
uniformly for $x \in \ER$, for some $ \rho < p < \infty$, and where $\rho$ is the positive root of
$\frac{1}{2}\frac{\rho -1}{\rho +1}=\frac{1}{\rho}$.
Then, there is a $\delta > 0$ such that for all
 $\phi_- \in W_{2,2} \cap  W_{1,1+\frac{1}{p}} $ with 
$\|\phi_-\|_{W_{2,2}}+ \|\phi_-\|_{W_{1,1+\frac{1}{p}} } \leq \delta$ there is a 
unique solution, $u$, to (\ref{3.1})
such that $u \in C(\ER, W_{1,2}) \cap M$ and,
\beq
\lim_{t \rightarrow -\infty}\|u(t)- e^{-itH} \phi_- \|_{W_{1,2}}=0.
\label{3.12}
\ene
Moreover, there is a unique $\phi_+ \in W_{1,2} $ such that
\beq
\lim_{t \rightarrow  \infty}\|u(t)- e^{-itH} \phi_+ \|_{W_{1,2}}=0.
\label{3.13}
\ene
Furthermore, $e^{-itH} \phi_{\pm} \in M$ and
\beq
\left\| u- e^{-itH} \phi_{\pm} \right\|_M \leq C \left\|e^{-itH} \phi_{\pm}
\right\|_M^p,
\label{3.14}
\ene
\beq
\left\|\phi_+ - \phi_- \right\|_{W_{1,2}} \leq C \left[\left\|\phi_- 
\right\|_{W_{2,2}} +\left\|\phi_- \right\|_{W_{1, 1+\frac{1}{p}}}\right]^p.
\label{3.15}
\ene
The scattering operator, $S_{V_0} : \phi_- \hookrightarrow \phi_+$ is injective on
$ W_{1,1+\frac{1}{p}} \cap  W_{2,2}$.
\end{theorem}
 Observe that, $ \rho \approx 3.56$. Remark that we do not  to  restrict $F$ in such a way that 
energy is conserved.

To reconstruct the potential, $V_0$, we introduce below the scattering operator associated with 
asymptotic states that are solutions to the linear Schr\"{o}dinger equation 
with potential zero:
\beq
S:= W_+^{\ast}\, S_{V_0}\, W_-.
\label{3.16}
\ene

\begin{theorem}
Suppose that the assumptions of Theorem 3.1 are satisfied. Then for every
$\phi \in W_{2,2} \cap  W_{1,1+\frac{1}{p}}$
\beq 
\left.\frac{d}{d \epsilon} \,S(\epsilon \phi)\right|_{\epsilon=0}= S_L \phi,
\label{3.17}
\ene
where the derivative on the left-hand side of (\ref{3.17}) exists in the sense of
strong convergence  in $W_{1,2} \cap W_{1,p+1}$. 
\end{theorem} 
\begin{corollary}
Under the conditions of Theorem 3.1 the scattering operator, $S$, 
determines uniquely 
the potential $V_0$.
\end{corollary}
In the case where $F(x,u)= \sum_{j=1}^{\infty} V_j(x) |u|^{2(j_0+j)}u$, with fixed integer
$j_0$, we can also
reconstruct the $V_j,\,\,\, j=1,2, \cdots$. 
\begin{lemma}
Suppose that the conditions of Theorem 3.1 are satisfied, and moreover, that 
$F(x,u)= \sum_{j=1}^{\infty} V_j(x) |u|^{2(j_0+j)} u,$\, where $j_0$ is an 
integer such that, $j_0 \geq (p-3)/2$,
\, for $|u| \leq \eta$, for 
some $\eta > 0$, and where $V_j \in W_{1, \infty}$ with $\|V_j\|_{W_{1, \infty}} \leq
M^j, j=1,2, \cdots$, for some positive constant $M$. Then, for 
any $\phi \in W_{2,2} \cap  W_{1,1+\frac{1}{p}} $ there is an 
$\epsilon_0 > 0$ such that for all $ 0 < \epsilon  < \epsilon_0$:
\beq
i \left( (S_{V_0}-I )(\epsilon \phi),\, \phi \right)_{L^2}=
\sum_{j=1}^{\infty}\epsilon^{2(j_0+j)+1}\left[ \int \int \, dt\, dx \, V_j(x) 
\left| e^{-itH}\phi \right|^{2(j_0+j+1)} +Q_j\right],
\label{3.18}
\ene
where $Q_1=0$ and $Q_j, j > 1$, depends only on $\phi$ and on $V_k$ with $k< j$.
Moreover, for any  $\acute{x} \in \ER$, and any
$ \lambda > 0$,  we denote, $\phi_{\lambda}(x):= 
\phi ( \lambda (x- \acute{x}))$.
Then, if $ \phi \neq 0$:
\beq
V_j(\acute{x})=\frac{\lim_{\lambda \rightarrow \infty}\lambda^3 \int \int\, dt \,
dx \, V_j(x) \left|e^{-itH} \phi_{\lambda} \right|^{2(j_0+j+1)}}{\int \int\, dt \,
dx \,  \left|e^{-itH_0} \phi \right|^{2(j_0+j+1)}}.
\label{3.19}
\ene
\end{lemma}
\begin{corollary}
Under the conditions of Lemma 3.4 the scattering operator, S, determines uniquely the
potentials $V_j, j=0,1,\cdots$.
\end{corollary}
The method to reconstruct the potentials $V_j, j=0,1,\cdots$, is as follows. First
we obtain $S_L$ from $S$ using (\ref{3.17}). By any standard method for inverse 
scattering for the linear Schr\"{o}dinger equation on the line we reconstruct $V_0$
(recall that $H$ has no eigenvalues).
We then reconstruct $S_{V_0}$ from $S$ using (\ref{3.16}). Finally (\ref{3.18}) and
(\ref{3.19}) give us, recursively, $V_j, j=1,2, \cdots$.

Theorems 3.1, 3.2, Lemma 3.4 and Corollaries 3.3, 3.5 are proven in \cite{we7} where
also a discussion of the literature is given. We give below an idea of the proofs.

In Theorem 1.1 of \cite{40} we proved that $\w$ and $\a$ are bounded
operators on $ W_{k,p},k=0,1,\, 1 < p < \infty$.  By Theorem 3 in page 135 of 
\cite{9},
\beq
\|{\cal F}^{-1} (1+q^2)^{k/2}({\cal F} f)(q)\|_{L^p},
\label{4.2}
\ene
is a norm that is equivalent to the norm of $W_{k,p}, 1 < p < \infty$. 
In (\ref{4.2}) ${\cal F}$ denotes the 
Fourier transform. Then, by the continuity of the $\w$ and $\a$ on $W_{k,p}$
(see Corollary 1.2 of \cite{40})
\beq
\| (I+H)^{k/2}\,f \|_{L^p},
\label{4.3}
\ene
defines a norm that is equivalent to the norm of $W_{k,p}, k=0,1, 1 < p < \infty$.

Condition (\ref{3.10b}) and Theorem 2.7.1 in page 35 of \cite{s} imply that, 
$D(H)=D(H_0)= W_{2,2}$, and that the following norm is equivalent to the norm of
$W_{2,2}$:
\beq
\left\| (H+I) \phi \right\|_{L^2}.
\label{4.9b}
\ene
The weight $(I+H)^{k/2}, k=1,2$ has the advantage that it commutes with 
$e^{-it H}$ and moreover, in the case $p=2$ the equivalent norm is invariant under
the time evolution given by $e^{-it H}$. We will  use these equivalences without further comments.
In particular, it 
follows from (\ref{4.3}) that estimate (\ref{p}) holds in the norm on    
 ${\cal B}\left( W_{1,p} ,W_{1,\acute{p}}\right),\,\, 1 < p \leq 2$.

 By Sobolev's imbedding theorem  \cite{8}, $W_{1,2}$ is continuously imbedded 
in $L^{\infty}$. It follows that $F$ is locally Lipschitz continuous on $W_{1,2}$. 
 Then, by 
standard arguments, $u \in C(\ER ,W_{1,2}) \cap M $ is a solution 
to (\ref{3.1}) with $\lim_{t \rightarrow - \infty} \|u(t) -
 e^{-itH} \phi \|_{\h}=0$, for some $ \phi \in W_{1, 2}$, if and only if
$u$ is a solution to the following integral equation:
\beq
u= e^{-itH} \phi + \frac{1}{i} \int_{-\infty}^t e^{-i(t -\tau )H} F(x,u(\tau ))\, d \tau.
\label{4.4}
\ene
As we prove below the integral in the right-hand side of  (\ref{4.4}) 
converges absolutely in ${\h}$ and in $M$. For $u \in M$ we denote
\beq
{\cal Q}u(t):=\frac{1}{i}  \int_{-\infty}^t e^{-i(t-\tau )H} F(x,u(\tau ))\, d \tau .
\label{4.5}
\ene 
It follows from  (\ref{p}), 
and since$W_{1, p+1}$  is continuously imbedded in $L^{\infty}$, that

\beq
   \left\| {\cal Q}u(t) - {\cal Q}v(t) \right\|_{\X} \leq C \,
(1+|t|)^{-d}( \|u\|_M + \|v \|_M )^{p-1}
\|u-v\|_M,
\label{4.6}
\ene
where we used that $p\, d > 1$. The  constants $C$ in (\ref{4.6})
can be taken uniform in closed balls in $M$.
By (\ref{4.6}) with $v(t)=0$ :
$$
\left\|{\cal Q}u(t) \right\|_{\h}^2\leq C \Re \int_{-\infty}^t \,d \tau \, 
\left(\sqrt{I+H} F(x, u(\tau)), \sqrt{I+H}{\cal Q}u(\tau) \right)_{L^2} \leq \,C
 \int_{-\infty}^t d \tau \, 
\|F(x, u)(\tau )\|_{W_{1,{1+1/p}}}\times
$$
$$
(1+ |\tau |)^{-d}\, \|u\|_M^p 
\leq C
\int_{-\infty}^t\, d \tau \, \|u\|_{W_{1,p+1}}^p\, \, (1+|\tau |)^{-d}
 \|u\|^p_M \leq
C \int_{-\infty}^t \,d \tau \, (1+|\tau |)^{-d(p+1)}\, \|u\|^{2p}_M 
$$
\beq
\leq C
 (1+\max[0,-t ])^{-(d+dp-1)}\, \|u\|^{2p}_M .
\label{4.7}
\ene
By (\ref{4.6}) with $v(t)=0$, the integral in the right-hand side of (\ref{4.4})
 converges in $M$ 
and by (\ref{4.7}) the  converge holds also in $\h$.

By (\ref{4.9b}) and  Sobolev's imbedding theorem,

\beq
\|e^{-itH} \phi_- \|_{W_{1,p+1}} \leq C \| e^{-itH}\phi_- \|_{W_{2,2}}\leq C
\left\| (H+I) e^{-itH} \phi_- \right\|_{L^2}= C 
\left\| (H+I)  \phi_- \right\|_{L^2}\leq C \left\| \phi_- \right\|_{W_{2,2}}.
\label{4.10}
\ene
Then, (\ref{p}) and (\ref{4.10}) imply that,

\beq
\left\|e^{-itH} \phi_- \right\|_M \leq C \left[ \left\| \phi_-\right\|_{W_{2,2}}+
\| \phi_- \|_{W_{1,1+\frac{1}{p}}}\right].
\label{4.11}
\ene  
For $R > 0$ let us denote: $ M_{R}:=\{ u \in M: \|u\|_M \leq R\}$. Let us take $R$ so
 small that $C (2 R)^{p-1} \leq 1/2$, with $C$ as in (\ref{4.6}), and  $ \delta > 0$ so
 small that $C \delta \leq R/4$, with $C$ as in (\ref{4.11}).
 It follows from  (\ref{4.6})
and (\ref{4.11}) that the map $ u \hookrightarrow
e^{-itH} \phi_- + {\cal Q} \, u$ is a contraction from $M_{R}$ into $M_{R}$
for all $\phi_- \in W_{2,2} \cap W_{1,1+\frac{1}{p}}$ with 
$\| \phi_- \|_{W_{2,2}}+\|\phi_-\|_{W_{1,1+\frac{1}{p}}} \leq \delta$. The 
contraction mapping theorem implies that there is an unique solution to 
(\ref{4.4}) in $M_{R}$. This is the solution $u(t)$ of Theorem 3.1. Moreover,
\beq
\left\|u\right\|_M \leq  \left\|e^{-itH} \phi_- \right\|_M +\frac{1}{2} 
\left\|u\right\|_M.
\label{4.13}
\ene
Then,
\beq
\left\|u\right\|_M \leq C \left\|e^{-itH} \phi_- \right\|_M.
\label{4.14}
\ene

We define:
\beq
\phi_+ =\phi_- + \frac{1}{i} \int_{-\infty}^{\infty} e^{i\tau H} F(x, u(\tau ))\, d \tau.
\label{4.15}
\ene
For further 
details on the proof of Theorem 3.1 see \cite{we7}.

\noindent {\it Proof of Theorem 3.2 :} Since, $S(0)=0$,
and $W_{\pm}$ are bounded on $W_{2,2} \cap W_{1,1+\frac{1}{p}}$ \cite{40},
  it is enough  to prove that
\beq
s-\lim_{\epsilon \rightarrow 0} \frac{1}{\epsilon}\,
(S_{V_0}(\epsilon \phi)-\epsilon \phi )
=0.
 \label{4.22}
\ene
By  (\ref{4.11}) and (\ref{4.14}) with $\phi_-$ replaced by $\epsilon \phi$:
\beq
\|u\|_M  \leq C |\epsilon| \left[ \left\| \phi \right\|_{W_{2,2}}+
\| \phi \|_{W_{1,1+\frac{1}{p}}}\right].
\label{4.23}
\ene

Using and (\ref{p}) and (\ref{4.15}) we obtain that, 
$$
   \left\|S_{V_0}(\epsilon \phi)-\epsilon \phi \right\|^2_{W_{1,2}}
     \leq C  \int_{-\infty}^{\infty} \,d \tau \, 
\left(\sqrt{I+H} F(x, u(\tau)), \sqrt{I+H} \int_{-\infty}^{\infty} d \rho \, e^{-i(\tau - \rho )H}
F(x, u(\rho )) \right)_{L^2} 
$$
$$
\leq \,C
 \int_{-\infty}^{\infty} d \tau \, 
\|F(x, u)(\tau )\|_{W_{1,{1+1/p}}}\times
$$
$$
(1+ |\tau |)^{-d}\, \|u\|_M^p 
\leq C
\int_{-\infty}^{\infty}\, d \tau \, \|u\|_{W_{1,p+1}}^p\, \, (1+|\tau |)^{-d}
 \|u\|^p_M \leq
C \int_{-\infty}^{\infty} \,d \tau \, (1+|\tau |)^{-d(p+1)}\, \|u\|^{2p}_M 
$$
\beq
\leq C \|u\|^{2p}_M .
\label{4.23b}
\ene
Equation (\ref{4.22})  follows from (\ref{4.23}) and (\ref{4.23b}).

\noindent {\it Proof of Corollary 3.3:}\,By Theorem 3.2 $S$ determines uniquely 
$S_L$. From $S_L$ we 
get the
reflection coefficients for linear Schr\"odinger scattering on  the line. As $H$ has
 no bound states we 
uniquely reconstruct $V_0$ from one of the reflection coefficients by using any  
method for inverse scattering on the line.

\noindent {\it Proof of Lemma 3.4 :}\,\, By the contraction mapping theorem, 
\beq
u(t)= e^{-itH}\epsilon  \phi+ \sum_{n=1}^{\infty} {\cal Q}^n e^{-itH}\epsilon 
\phi.
\label{4.24}
\ene
Equation (\ref{3.18}) follows from (\ref{4.15}) and (\ref{4.24}).
By Sobolev's imbedding theorem \cite{8}, $W_{2,2} \subset L^{q},\, 2 \leq q \leq
\infty$. Then, estimating as in (\ref{4.10}) we prove that,
$\left\|e^{-itH} \phi \right\|_{L^q} \leq C_q \left\|
e^{-itH} \phi \right\|_{W_{2,2}} \leq  C_q\left\|
 \phi \right\|_{W_{2,2}}, 2 \leq  q \leq \infty $ , and as $ 2(j_0 +j +1) \geq  p+1$  it follows from (\ref{p}) 
that:
\beq
\int \int\, dt \,
dx \,  \left|e^{-itH} \phi  \right|^{2(j_0+j+1)}\leq \left\|e^{-itH}\phi 
\right\|_{L^{\infty}}^{2(j_0 +j +1)- p-1}    \int\, dt \,
dx \,  \left|e^{-itH} \phi  \right|^{p+1}  < \infty,\,\,\, j=1,2, \cdots.
\label{4.25}
\ene                               
  For 
$ \lambda > 0$ and $ \acute{x} \in \ER$ we denote
by $H_{\lambda}$ the following self-adjoint operator in $L^2$:
\beq
H_{\lambda}:= H_0 + V_{\lambda}(x),\, \hbox{where}\, V_{\lambda}(x)= 
\frac{1}{\lambda^2}V_0\left(\frac{x}{\lambda}+ \acute{x}\right).
\label{3.17b}
\ene
Since $H$ has no eigenvalues, we have that $H_{\lambda}$ has no eigenvalues, i.e.,
$ H_{\lambda} > 0$.
It follows from  (\ref{3.10b}) and from Theorem 2.7.1 on page 35 of \cite{s}
that
\beq
C_1 \left\| \phi \right\|_{W_{2,2}} \leq \left\| (H_{\lambda}+I) \phi \right\|_{L^2}
 \leq C_2 \left\| \phi \right\|_{W_{2,2}},
\label{4.26}
\ene
for some constants $C_1, C_2$. Moreover, since $N(V_{\lambda}) \leq 
\frac{1}{\lambda^3}\,N (V_0),\,\,\, \lambda \geq 1 $, the proof of 
Theorem 2.7.1 on page 35 of \cite{s} implies that
we can take fixed $C_1$ and $C_2$  for all $ \lambda \geq 1$.
 To prove (\ref{3.19}) we denote: $\tilde{t}:= \lambda^2 t$ and 
$\tilde{x}:= \lambda (x-\acute{x})$. Then, we observe that,
\beq
\left( e^{-i\tilde{t} H_{\lambda}}  \phi\right)(\tilde{x})=
\left( e^{-itH} \phi_{\lambda}\right)(x).
\label{4.27}
\ene
This can be seen as follows,
\beq
i \frac{\partial}{\partial t} \left( e^{-i\tilde{t} H_{\lambda}}  \phi\right)=
H \left( e^{-i\tilde{t} H_{\lambda}}  \phi\right),\,\,\, \hbox{and}\,\,\, 
\left( e^{-i\tilde{t} H_{\lambda}}  \phi\right)\left.\right|_{t=0}= \phi_{\lambda}.
\label{4.27b}
\ene
Since the solution to the linear Schr\"{o}dinger equation is unique, (\ref{4.27}) is proved.  
It follows from (\ref{4.27}) that,
\beq
I_j:= \lambda^3 \int \int\, dt \,
dx \, V_j(x) \left|e^{-itH} \phi_{\lambda} \right|^{2(j_0+j+1)}=
\int \int\, d\tilde{t} \,
d\tilde{x} \, V_j(\frac{\tilde{x}}{\lambda}+\acute{x}) 
\left|e^{-i\tilde{t}H_{\lambda}}   \phi \right|^{2(j_0+j+1)}
 (\tilde{x}).
\label{4.28}
\ene
By  (\ref{4.26})

\beq
s- \lim_{\lambda \rightarrow \infty}
 e^{-i\tilde{t}H_{\lambda}}  \phi =  
 e^{-i\tilde{t}H_0}  \phi,
\label{4.29}
\ene
where the limit exists in the strong topology on $W_{2,2}$. By Sobolev's imbedding
theorem, the limit in (\ref{4.29}) also exists in the strong topology on $L^q, \,
2 \leq q\leq \infty$, and moreover,
\beq
\left\|e^{-i\tilde{t}H_{\lambda}}  \phi\right\|_{L^q} \leq
C_q \left\| \phi \right\|_{W_{2,2}}, \,\,\,2 \leq q \leq \infty,\,\,\, \lambda \geq 1.
\label{4.30}
\ene
 Furthermore, by  (\ref{p}) and(\ref{4.27})

\beq 
\left\|e^{-i\tilde{t} H_{\lambda}}\phi \right\| _{L^{p+1}}^{p+1}= \lambda  \left\|e^{-i t H}\phi_{\lambda} 
\right\| _{L^{p+1}}^{p+1}
\leq C  \frac{1}{ t^{d(p+1)}} \lambda  \left\| \phi_{\lambda} \right\|_{L^{1+1/p}}^{p+1} = C
\frac{1}{ \tilde{t}^{d(p+1)}} \left\|\phi \right\|_{L^{1+1/p}}^{p+1},
\label{4.31}
\ene
with $d:= \frac{1}{2}\frac{p-1}{p+1}$. Equation (\ref{3.19}) follows from 
(\ref{4.28}), (\ref{4.29}), (\ref{4.30}),
(\ref{4.31}) and the dominated convergence theorem, observing that $2(j_0+j+1) \geq
p+1$, that $d (p+1) > 1$ and that $V_j$ is continuous.

\noindent {\it Proof of Corollary 3.5:}\,\, By Corollary 3.3, $S$ determines 
uniquely $V_0$. Then the 
wave operators, $W_{\pm}$, are uniquely determined, and by (\ref{3.16}), $S$ 
determines uniquely $S_{V_0}$. Finally by (\ref{3.18}) and (\ref{3.19}) $S_{V_0}$ 
determines
uniquely $V_j, j=1,2, \cdots$.


\begin{thebibliography}{99}
 \bibitem{8}  Adams, R.A., Sobolev Spaces, {\sl Academic Press}, New York, 1975.

\bibitem{ab} Aharonov Y., Bohm D., Significance of electromagnetic potentials in 
the quantum
theory, {\sl  Phys. Rev.} 115: 485-491 (1959).


\bibitem{a1} Arians, S., Geometric Approach to inverse scattering for the 
Schr\"{o}dinger equation with magnetic and electric potentials. {\sl J. Math. Phys.}
38: 2761 - 2773, 1997.

\bibitem{a2} Arians, S., Geometric approach to inverse scattering for hydrogen
like systems in a homogeneous magnetic field. {\sl J. Math. Phys.} 39: 1730 - 1743,
1998.

\bibitem{a3} Arians, S., Inverse Streutheorie f\"{u}r die 
Schr\"{o}dingergleichung mit Magnetfeld. {\sl Dissertation RWTH Aachen
; Logos-Verlag},
Berlin, 1998.

\bibitem {DT} Deift, P. and  Trubowitz, E., Inverse scattering 
on the line. Commun. Pure Appl.
Math. XXXII: 121 - 251, 1979.

\bibitem{ew1} Enss, V., Weder, R., Inverse potential scattering: A geometrical
 approach. In: {\sl Mathematical Quantum Theory II: Schr\"{o}dinger Operators, 
CRM Proc. Lecture Notes} vol 8, pp. 151 - 162, {\sl American Mathematical Society}, 
Providence, 1995 ( Proceedings Vancouver 1993).

\bibitem{ew2}Enss, V., Weder, R., Uniqueness and reconstruction formulae for inverse N-particle
scattering. In: {\sl Differential Equations and Mathematical Physics},
pp. 55-66, {\sl International Press}, Boston, 1995 (Proceedings Birmingham AL 1994).

\bibitem{ew3} Enss, V., Weder, R., The geometrical approach to multidimensional inverse 
scattering. {\sl J. Math. Phys.} 36: 3902 - 3921, 1995.

\bibitem{ew4}Enss, V., Weder, R., Inverse two-cluster 
scattering. {\sl Inverse Problems} 12: 
409 - 418, 1996.

\bibitem{he} Helgason, S., Groups and Geometrical Analysis , {\sl Academic
Press}, Orlando, 1984.


\bibitem{it} Ito, H. T., An inverse scattering problem for the Dirac equation 
with time-dependent electromagnetic fields. {\sl Pub. Res. Inst. Math. Sci.} 34: 355
 - 381, 1998.

\bibitem{j1} Jung, W., Der geometrische Ansatz zur inversen Streutheorie bei der 
Dirac-Gleichung, {\sl Diplomarbeit RWTH Aachen}, 1996.

\bibitem{j2} Jung, W., Geometrical approach to inverse scattering for the Dirac equation.
{\sl J. Math. Phys.} 38: 39 - 48, 1997.

\bibitem{ka} Kato T., Perturbation Theory of Linear Operators, Second Edition, {\sl Springer-Verlag},
Berlin, 1978.



\bibitem{na} Natterer F., The Mathematics of Computerized Tomography, {\sl B. Teubner,
 J. Wiley \& Sons}, Stuttgart and Chichester, 1989.

\bibitem{ni} Nicoleau F., An inverse scattering problem with the Aharonov-Bohm 
effect,  {\sl J. Math. 
Phys.} 41: 5223-5237 (2000).



\bibitem{rs} Reed, M., Simon, B., Methods of Modern Mathematical Physics
III. Scattering Theory, {\sl Academic Press}, New York, 1979.

\bibitem{s}  Schechter, M.,  Operator Methods in Quantum Mechanics, 
{\sl North Holland}, New York, 1981.

\bibitem{9}  Stein, E.M., Singular Integrals and Differentiability Properties
 of Functions,
 {\sl Princeton Univ. Press}, Princeton, New Jersey, 1970.


\bibitem{we1} Weder, R., Characterization of the scattering data in 
multidimensional inverse scattering theory. {\sl Inverse Problems} 7: 461 - 489,
 1991. 

\bibitem{we2} Weder, R., Multidimensional inverse scattering in an electric
 field. {\sl J. Funct. Anal.} 139: 441 - 465, 1996.

\bibitem{we3} Weder, R., Inverse scattering for N-body systems with 
time-dependent potentials. In: {\sl Inverse Problems of Wave Propagation and 
Difraction}, pp. 27 - 46, { \sl Lecture Notes in Physics, Springer-Verlag}, Berlin,
1997 (Proceedings Aix les Bains 1996).

\bibitem{we4} Weder, R., Inverse scattering for the nonlinear Schr\"{o}dinger 
equation. {\sl Comm. Partial Differential Equations} 22: 2089 - 2103, 1997.

\bibitem{40} Weder, R., The $W_{k,p}$-continuity of the Schr\"{o}dinger wave 
operators on the line. {\sl Comm. Math. Phys.} 208: 507-520, 1999.

\bibitem{we5} Weder, R., $L^p-L^{\acute{p}}$ estimates for the Schr\"{o}dinger
equation on the line and inverse scattering for the nonlinear Schr\"{o}dinger 
equation with a potential. {\sl J. Funct. Anal.} 170: 37 - 68, 2000.

\bibitem{we6} Weder, R., Uniqueness of inverse scattering for the nonlinear
Schr\"{o}dinger equation  and reconstruction of the potential and the nolinearity.
In: {\sl Proceedings of the Fifth International Conference on Mathematical and 
Numerical Aspects of Wave Propagation}, pp. 631 - 634, {\sl SIAM Proceedings
Series, Society for Industrial and Applied Mathematics}, Philadelphia, 2000
(Proceedings Santiago de Compostela 2000).

\bibitem{we12}  Weder, R., $L^p-L^{\acute{p}}$ estimates for the Schr\"{o}dinger
equation and inverse scattering. In: {\sl Differential Equations and Mathematical 
Physics}, pp. 435 - 448, {\sl American Mathematical Society and International Press},
Providence, 2000 ( Proceedings Birmingham AL 1999). 

\bibitem{we7} Weder, R., Inverse scattering for the nonlinear Schr\"{o}dinger
equation. Reconstruction of the potential and the nonlinearity.
 {\sl Mathematical Methods in the Applied Sciences} 24: 245-254, 2001.

 \bibitem{we8} Weder, R., Inverse scattering for the nonlinear Schr\"{o}dinger
equation II. Reconstruction of the potential and the nolinearity in the 
multidimensional case. {\sl  Proceedings of the
American Mathematical Society} 129: 3637-3645, 2001.

\bibitem{we9} Weder, R., Inverse Scattering on the line for the 
nonlinear Klein-Gordon equation with a potential. {\sl J. Math. Anal. Appl.}:
252, 102-123, 2000.

\bibitem{we10} Weder, R., Multidimensional inverse scatering for the nonlinear 
Klein-Gordon equation with a potential.  Preprint 2001. To appear in Journal of Differential
Equations.  

\bibitem{we11} Weder R., Aharonov-Bohm effect and time-dependent inverse scattering
theory. Preprint 2001, http://rene.ma.utexas.edu/mp\_ arc-bin/mpa?yn=01-318.

\bibitem{wei} Weidmann, J., Spectral Theory of Ordinary Differential Operators, {\sl Lecture
Notes in Mathematics  1258, Springer-Verlag}, Berlin 1987.
\end{thebibliography}
\end{document}